\documentclass[aps,prl,twocolumn,showpacs,showkey,square,numbers,amssymb,amsmath,nobibnotes,footinbib]{revtex4}
\usepackage{bm}
\usepackage{amsmath}
\usepackage{amsfonts}
\usepackage{amssymb}   
\usepackage{verbatim}
\usepackage{times}
\usepackage{graphicx}
\usepackage{dsfont}
\usepackage{color}

\newcommand{\barr}{\begin{eqnarray}}
\newcommand{\earr}{\end{eqnarray}}

\newcommand{\beq}{\begin{equation}}
\newcommand{\eeq}{\end{equation}}
\newcommand{\be}{\begin{equation}}
\newcommand{\ee}{\end{equation}}

\newcommand{\dd}{\mathrm{d}}

\begin{document}


\title{Phase transitions and edge scaling of number variance in Gaussian random matrices}



\author{Ricardo Marino, Satya N. Majumdar, Gr\'egory Schehr and Pierpaolo Vivo}
\affiliation{Laboratoire de Physique Th\'{e}orique et Mod\`{e}les
Statistiques (UMR 8626 du CNRS), Universit\'{e} Paris-Sud,
B\^{a}timent 100, 91405 Orsay Cedex, France}

\date{\today}


\begin{abstract}

We consider $N\times N$ Gaussian random matrices, whose average density of eigenvalues has the Wigner semi-circle form 
over $[-\sqrt{2},\sqrt{2}]$. For such matrices, using a Coulomb gas technique, we compute the large $N$ behavior of the probability $\mathcal{P}_{\scriptscriptstyle N,L}(N_L)$ that $N_L$ eigenvalues lie within the box $[-L,L]$. This probability scales as $\mathcal{P}_{\scriptscriptstyle N,L}(N_L=\kappa_L N)\approx\exp\left(-{\beta} N^2 \psi_L(\kappa_L)\right)$, where $\beta$ is the Dyson index of the ensemble and $\psi_L(\kappa_L)$ is a $\beta$-independent rate function that we compute exactly. We identify three regimes as $L$ is varied: (i) $\, N^{-1}\ll L<\sqrt{2}$ (bulk), (ii) $\ L\sim\sqrt{2}$ on a scale of $\mathcal{O}(N^{-{2}/{3}})$ (edge) and (iii) $\ L > \sqrt{2}$ (tail). We find a dramatic non-monotonic behavior of the number variance $V_N(L)$ as a function of $L$: after a logarithmic growth $\propto \ln (N L)$ in the bulk (when $L \sim {\cal O}(1/N)$), $V_N(L)$ decreases abruptly as $L$ approaches the edge of the semi-circle before it decays as a stretched exponential for $L > \sqrt{2}$. This ``drop-off'' of $V_N(L)$ at the edge is described by a scaling function $\tilde V_{\beta}$ which smoothly interpolates between the bulk  (i) and the tail (iii). For $\beta = 2$ we compute $\tilde V_2$ explicitly in terms of the Airy kernel. These analytical results, verified by numerical simulations, directly provide for $\beta=2$ the full statistics of particle-number fluctuations at zero temperature of 1d spinless fermions in a harmonic trap.

\end{abstract}


\maketitle

There has been enormous progress in the last decade on the experimental manipulation of cold atoms \cite{bloch,reviewfermi} that generated several
interesting theoretical questions concerning the interplay between quantum and statistical behaviors in many-body systems. These experiments are usually carried out in presence of optical laser traps that confine the particles in a limited region of space. In particular, one-dimensional systems, such as spinless fermions in presence of a harmonic trap, have played a crucial role in these recent developments \cite{reviewfermi,calabrese_prl,vicari_pra,vicari_pra2,vicari_pra3,eisler_prl}. One important observable that has been studied is the number of fermions $N_L$ in the ground state ($T=0$) within a given box $[-L,+L]$. The variance of $N_L$, denoted by $V_N(L)$, characterizes the quantum fluctuations in the ground state of this many-body system. This quantity has also been studied recently in a number of other quantum systems, including several lattice models of fermions \cite{eisler_racz,eisler_peschel}.

The variance $V_N(L)$ as a function of the box size $L$ turns out to be highly nontrivial even for the simplest possible many-body quantum system, namely 1d spinless fermions in a harmonic trap. In this case, it was numerically found that $V_N(L)$ has a rather rich non-monotonic dependence on $L$ -- it first increases with $L$ and then drops rather dramatically when $L$ exceeds some threshold value \cite{vicari_pra,eisler_prl}. Analytically deriving this dependence on $L$ is thus a challenging problem. In this Letter, exploiting a connection of this fermionic system at $T=0$ to the Gaussian Unitary Ensemble (GUE) of random matrices, we obtain, for large $N$,  this variance $V_N(L)$ exactly for arbitrary $L$, which explains its non-monotonic behavior. In addition, using a Coulomb gas technique for random matrices, we are able to calculate the full probability distribution of $N_L$ in the large $N$ limit.        

\begin{figure}[ht]
\centering
\includegraphics[width=\linewidth]{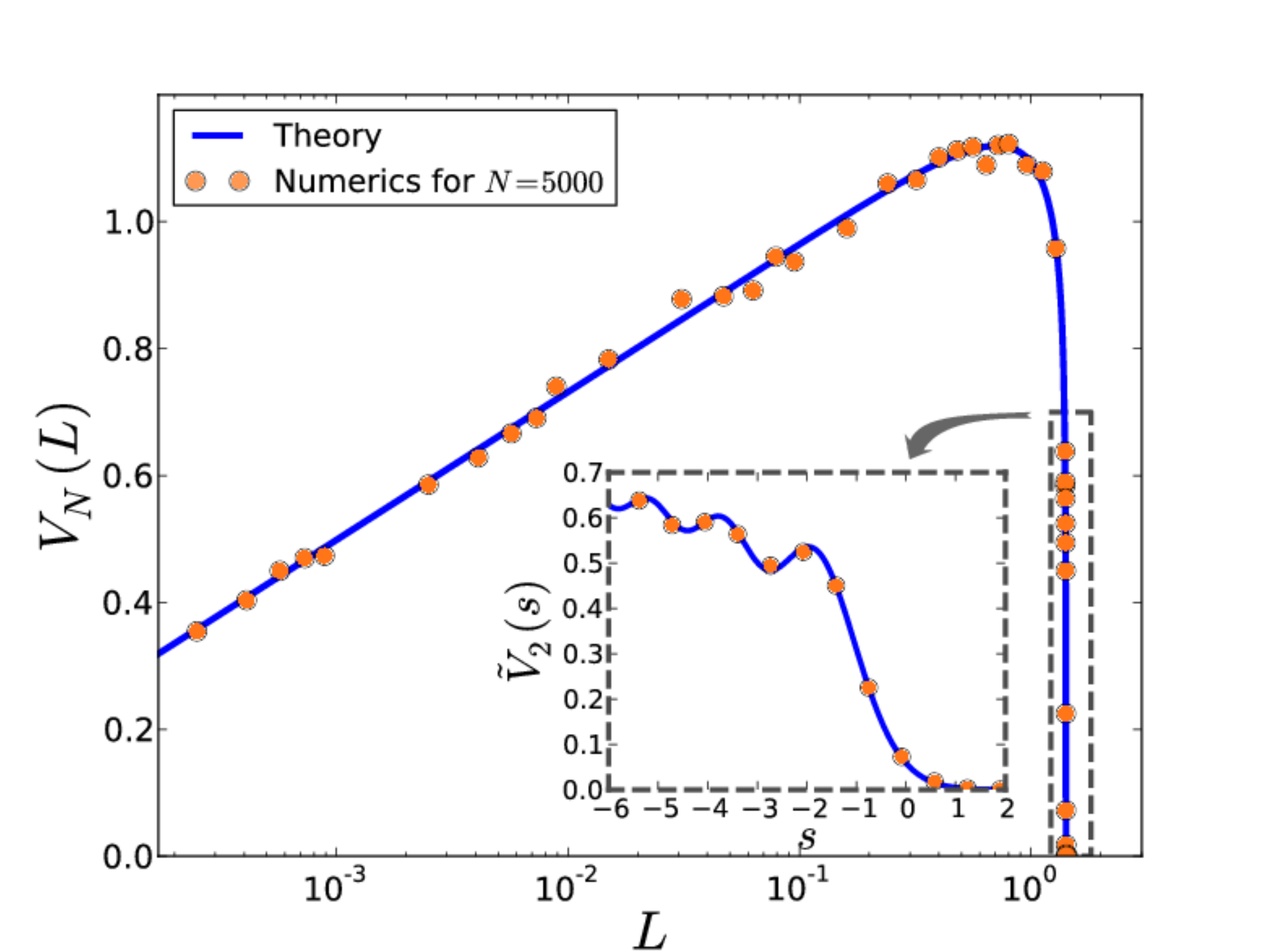} 
\caption{Number variance $V_N(L)$ as a function of $L$. Theoretical result Eq. \eqref{variance_summary} in solid blue line. {\bf Inset}: edge scaling behavior of the variance around $L\sim\sqrt{2}$ (described by the scaling function $\tilde{V}_2(s)$, in \eqref{variance_dble}, with $s=(L-\sqrt{2})\sqrt{2}N^{2/3}$) together with numerical simulations for $N=5000$ and $\beta=2$ (and averaged over 30000 matrices).}
\label{fig_variance}
\end{figure}

The ground state many-body wavefunction of $N$ 1d spinless fermions in a harmonic potential $U(x) = \frac{1}{2} m \omega^2 x^2$ is given by the Slater determinant $\Psi_0(\vec{x}) = \frac{1}{\sqrt{N!}}\det [\varphi_i(x_j)]$ where $\varphi_n(x)$ is the single particle harmonic oscillator wavefunction, $\varphi_n(x) \propto H_n(x) \mathrm{e}^{-x^2/2}$ (we have set $\hbar = m = \omega =1$), where $H_n(x)$ are Hermite polynomials. By explicitly evaluating this determinant, it is easy to see that
\begin{eqnarray}\label{eq_psi0}
|\Psi_0(\vec{x})|^2 = \frac{1}{Z_N} \mathrm{e}^{-\sum_{i=1}^N x_i^2} \prod_{j<k} (x_j - x_k)^2  \;, 
\end{eqnarray}
where $Z_N$ is the normalization constant. The $|\Psi_0(\vec{x})|^2$ can thus be interpreted as the joint probability distribution function (PDF) of the eigenvalues $x_1, x_2, \cdots, x_N$ of an $N \times N$ GUE matrix \cite{mehta,forrester}. The quantum operator corresponding to the number of fermions in a box $[-L,L]$ is denoted by $\hat N_L = \int_{-L}^{+L} \hat n(y) dy$ with $\hat n(y) = c^\dagger(y) c(y)$ where $c^\dagger(y)$ and $c(y)$ are the standard fermion creation and annihilation operators at $y$. Using (\ref{eq_psi0}), it is easy to see that the ground state expectation of the $k$-th moment of $\hat N_L$, $\langle 0|(\hat N_L)^k |0\rangle$ is then identical to the $k$-th moment of the number of eigenvalues of GUE in $[-L,+L]$ which we denote by  $N_L = \sum_{i=1}^N \mathds{1}_{[-L,L]}(x_i)$, where $\mathds{1}_{\mathcal{I}}(x)$ is the indicator function $=1$ if $x\in\mathcal{I}$ and zero otherwise. Hence studying the quantum fluctuations of $\hat N_L$ at $T=0$ reduces to studying the statistics of the classical observable $N_L = \sum_{i=1}^N \mathds{1}_{[-L,L]}(x_i)$ in the GUE random matrix.

The statistics of $N_L$ is actually interesting to study for a general Gaussian random matrix, not necessarily GUE. Here we consider the three standard Gaussian ensembles that are real symmetric (GOE, $\beta=1$), complex Hermitian (GUE, $\beta=2$), or quaternion self-dual (GSE, $\beta=4$), whose entries are independently drawn from a zero-mean Gaussian distribution and variance ${\cal O}(1/N)$ where $\beta=1,2,4$ is the Dyson index of the ensemble. In the following, we denote the positions $\{ x_i \}$'s of the fermions by $\{\lambda_i\}$'s to stick to the standard random matrix notations.  The joint PDF of the $N$ real eigenvalues is given by
\begin{eqnarray}
\mathcal{P}_\beta({\bm \lambda})= \frac{1}{Z_{N,\beta}} \mathrm{e}^{-\frac{\beta N}{2}\sum_{j=1}^N\lambda^2_j} \prod_{i>j}|\lambda_i-\lambda_j|^\beta\equiv \frac{\mathrm{e}^{-\beta E[\bm\lambda]}}{Z_{N,\beta}},\label{jpd}
\end{eqnarray}
where $Z_{N,\beta}$ is a normalization constant and $E[\bm\lambda]=(N/2)\sum_{i=1}^N\lambda_i^2-(1/2)\sum_{j\neq k}\ln|\lambda_j-\lambda_k|$. It is well known \cite{mehta} that, for large $N$, the average density $\rho_N(\lambda)$ of eigenvalues (normalized to unity) of a Gaussian random matrix approaches the celebrated Wigner's semicircle law on the \emph{compact} support $[-\sqrt{2},\sqrt{2}]$, $\rho_N(\lambda)\to\rho_{\text{sc}}(\lambda)=\pi^{-1}\sqrt{2-\lambda^2}$. The typical distance between eigenvalues is thus of order $\sim\mathcal{O}(1/N)$ near the center of the semi-circle.

The average $\langle N_L\rangle$ for large $N$ can be computed as $\langle N_L\rangle=N\int_{-L}^L\rho_{\mathrm{sc}}(x)\dd x=N \kappa_L^\star$, where $\kappa_L^\star=\left[L\sqrt{2-L^2} +2\ \mathrm{arcsin}\left(L/\sqrt{2}\right)\right]/\pi$. The variance $V_N(L)$  was also computed in \cite{dyson_62,dyson_mehta_62}, but only in the bulk limit, i.e. when $L \sim {\cal O}(1/N)$ (the box size is of the order of the interparticle spacing near the center). On this scale, setting $L= {\Delta}/{N}$, $V_N(L)$ was shown to grow logarithmically with $\Delta$,
$
V_N(L) \sim (2/\beta\pi^2)\ln(\Delta)
$, for $\Delta \gg 1$.
In contrast, 
numerical simulation in the fermionic system shows a non-monotonic behavior of $V_N(L)$ as $L$ increases beyond ${\cal O}(1/N)$. 
A natural question is then: can one calculate $V_N(L)$ for all $L$? In this Letter we indeed compute $V_N(L)$ for all $L$ in the large $N$ limit, which exhibits a striking ``drop-off effect'' near the semi-circular edge (see Fig. \ref{fig_variance}). Our method also allows us to compute, for arbitrary $L$,  the full PDF of $N_L$ for large $N$, which was known to be a Gaussian but only on the scale $L \sim {\cal O}(1/N)$  \cite{CL,FS,Soshnikov,forrester_lebowitz_13}.

Our results can be summarized as follows. We find that the number variance $V_N(L)$ for an interval $[-L,L]$ behaves as
\begin{eqnarray}\label{variance_summary}
V_N(L) \sim 
\begin{cases}
\frac{2}{\beta\pi^2}\ln \left(\! NL(2-L^2)^{\frac{3}{2}}\!\right)\! , \,N^{-1}\! \ll L < \sqrt{2}\\
\tilde V_{\beta}(s), \quad L = \sqrt{2} +\frac{s}{\sqrt{2}}N^{-\frac{2}{3}} \\
\exp{[-\beta N \phi(L)]} \;, \quad L > \sqrt{2} \ ,
\end{cases}
\end{eqnarray}
where the scaling function $\tilde{V}_\beta(s)$ is computed explicitly in~\eqref{variance_dble} for $\beta = 2$ -- its asymptotic behaviors for generic $\beta$ are given in (\ref{asympt_vtilde}) -- and the function $\phi(L)$ is given in~\eqref{phiL} (in this third regime $\sim$ stands for a logarithmic equivalent). In (\ref{variance_summary}), $L \gtrless \sqrt{2}$ means $|L-\sqrt{2}| \gg N^{-2/3}$.

\begin{figure}[h]
\centering
\includegraphics[width=\linewidth]{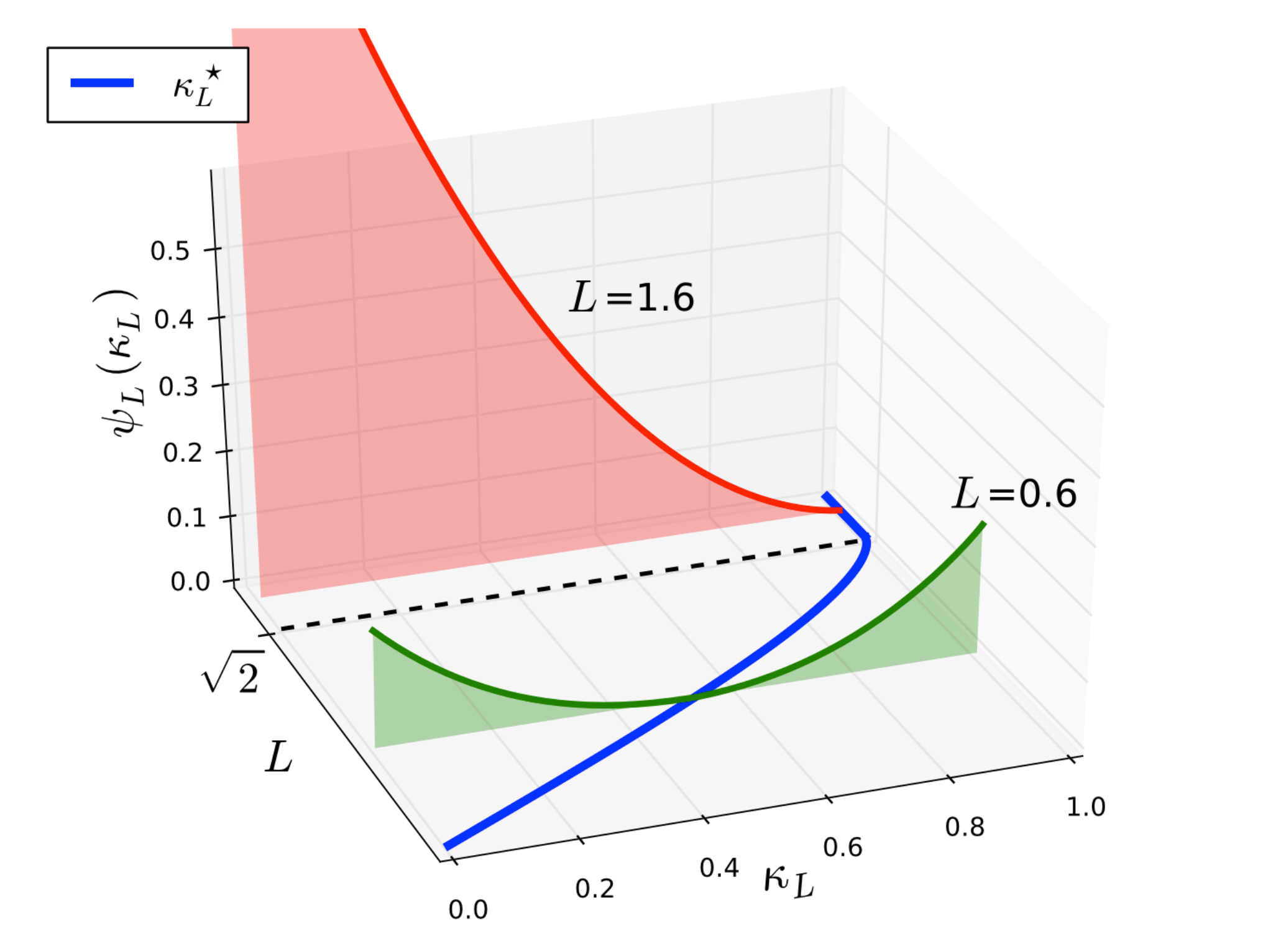} 
\caption{ Behavior of the rate function  $\psi_L(\kappa_L)$ as a function of $\kappa_L\in [0,1]$ for two different values of $L$: $L=0.6 < \sqrt{2}$ (green) and $L=1.6 > \sqrt{2}$ (red). The solid blue line in the plane $(L,\kappa_L)$ is the critical line $\kappa_L^\star$, where $\psi_L(\kappa_L)$ has a minimum (zero).}
\label{rate_3D}
\end{figure}

We thus identify three qualitatively different regimes depending on the value of $L$: (i) $\, N^{-1}\ll L< \sqrt{2}$ (bulk), (ii) $L \sim \sqrt{2}$,  on a scale of $\mathcal{O}(N^{-2/3})$ (edge) and (iii) $L > \sqrt{2}$ (tail). Moreover, we are able to obtain the \emph{full} probability distribution $\mathcal{P}_{\scriptscriptstyle N,L}(N_L)$ of $N_L$, for large $N$. Calling $\kappa_L=N_L/N$ the fraction of eigenvalues in $[-L,L]$ we obtain~\footnote{The symbol $\approx$ stands for a logarithmic equivalence, $\lim_{N\to\infty} -\ln \mathcal{P}_{\scriptscriptstyle N,L}(N_L=\kappa_L N)/\beta N^2=\psi_L(\kappa_L)$.}
\be\mathcal{P}_{\scriptscriptstyle N,L}(N_L=\kappa_L N)\approx\exp\left(-\beta N^2 \psi_L(\kappa_L)\right) \;,\label{generalrate}
\ee 
for $ 0\leq\kappa_L\leq 1$, where the rate function $\psi_L(\kappa_L)$ is $\beta$-independent and can be explicitly computed in terms of single integrals (see Eq. (56) in \cite{suppmat}). 
The rate function $\psi_L(\kappa_L)$ is convex and has a minimum (zero) at $\kappa_L=\kappa_L^\star$ (see Fig. \ref{rate_3D}). Thus the distribution of $N_L$ is peaked around $N_L=\kappa_L^\star N$ which is precisely its mean value $\langle N_L\rangle=\kappa_L^\star N$ for large $N$. This distribution has non-Gaussian tails and even near its peak in $\langle N_L\rangle$ it exhibits an anomalous quadratic behavior which is modulated here by a logarithmic singularity.

The probability distribution of $N_L$ -- considered here for simplicity as a continuous variable -- can be written by integrating the jpd \eqref{jpd} with a delta constraint $\delta\left(N_L-\sum_{i=1}^N \mathds{1}_{[-L,L]}(\lambda_i)\right)$, obtaining \cite{suppmat} 
\be
\mathcal{P}_{\scriptscriptstyle N,L}(N_L)\propto\int \prod_{i=1}^N \dd\lambda_i\int\frac{\dd \xi}{2\pi} \exp(-\beta E[{\bm \lambda};\xi,N_L]).\label{gibbs}
\ee 
The \emph{energy} $E[{\bm \lambda};\xi,N_L]$ of a configuration $\{{\bm \lambda}\}$ is then given by
$
E[{\bm \lambda};\xi,N_L]=E[\bm\lambda]+(\xi/2)\left(N_L-\sum_{i=1}^N \mathds{1}_{[-L,L]}(\lambda_i)\right)
$, where a Lagrange multiplier $\xi$ is introduced to take care of the delta constraint.  Written in this form, \eqref{gibbs} is just the grand-canonical partition function of a 2D fluid of charged particles (the eigenvalues) confined to a line. The system is in equilibrium at inverse temperature $\beta$ under competing interactions: a confining quadratic potential and a logarithmic all-to-all repulsion term. In addition, a fraction $\kappa_L=N_L/N$ of particles is constrained within the box $[-L,L]$. Introducing a normalized density of eigenvalues $\rho(\lambda)=N^{-1}\sum_{i=1}^N\delta(\lambda-\lambda_i)$, one converts the multiple integral in \eqref{gibbs} to a functional integral over $\rho$, which is then evaluated for large $N$ using a saddle point method. This 
procedure, originally introduced by Wigner and Dyson \cite{wigner_51,dyson_62}, has been successfully employed in recent works on the top eigenvalue of Gaussian and Wishart matrices \cite{dean_majumdar_06,dean_majumdar_08,vivo_majumdar_bohigas_07,majumdar_vergassola_09,majumdar_schehr_14}, conductance fluctuations in mesoscopic systems \cite{vivo_majumdar_bohigas_08} or bipartite entanglement of quantum systems \cite{facchi_marzolino_parisi_pascazio_scardicchio_08,nadal_majumdar_vergassola_10,nadal_majumdar_vergassola_11}. For the present problem, a similar Coulomb gas analysis was performed in Ref. \cite{FS} but it was restricted to the narrow regime $L \sim {\cal O}(1/N)$. Here we explore instead the full range $L \gg{N^{-1}}$ up to the edge, $L \sim \sqrt{2}$, and beyond. Skipping details \cite{suppmat}, the resulting expression for $\mathcal{P}_{\scriptscriptstyle N,L}(N_L=\kappa_L N)$ reads to leading order for large~$N$
\begin{eqnarray}
\mathcal{P}_{\scriptscriptstyle N,L}(N_L=\kappa_L N)\propto \int \mathcal{D}[\rho]\dd\xi\dd\eta\exp\left(-\frac{\beta}{2}N^2 S[\rho]\right),\label{resulting}
\end{eqnarray}
where the action (depending on $L$ and $\kappa_L=N_L/N$) is
\begin{align}
\nonumber S[\rho] &=\int \dd x x^2 \rho(x) - \iint \dd x \dd x^\prime \rho(x)\rho(x^\prime)\ln|x-x'| \\
&+ \eta\left(\int \dd x\rho(x)-1\right)+\xi\left(\int_{-L}^{L}\dd x \rho(x)-\kappa_L\right).
\label{eq:hamilt_1}
\end{align}
Here $\eta$ is another Lagrange multiplier enforcing the normalization of the density, and the first integrals run over $(-\infty,\infty)$.
We now evaluate the integral \eqref{resulting} for large $N$ with a saddle point method. Differentiating \eqref{eq:hamilt_1} functionally with respect to $\rho$ and then with respect to $x$, we obtain a singular integral equation for $\rho^\star(x)$ (depending on $L$ and $\kappa_L$)
\be
x = \mathrm{Pr}\int \frac{\rho^\star(x^\prime)}{x-x^\prime}\dd x^\prime,\quad x\in\mathrm{supp}(\rho^\star)\text{ and }x\neq \pm L,\label{eq:cauchy}
\ee
\begin{figure}
\centering
\includegraphics[width=\linewidth]{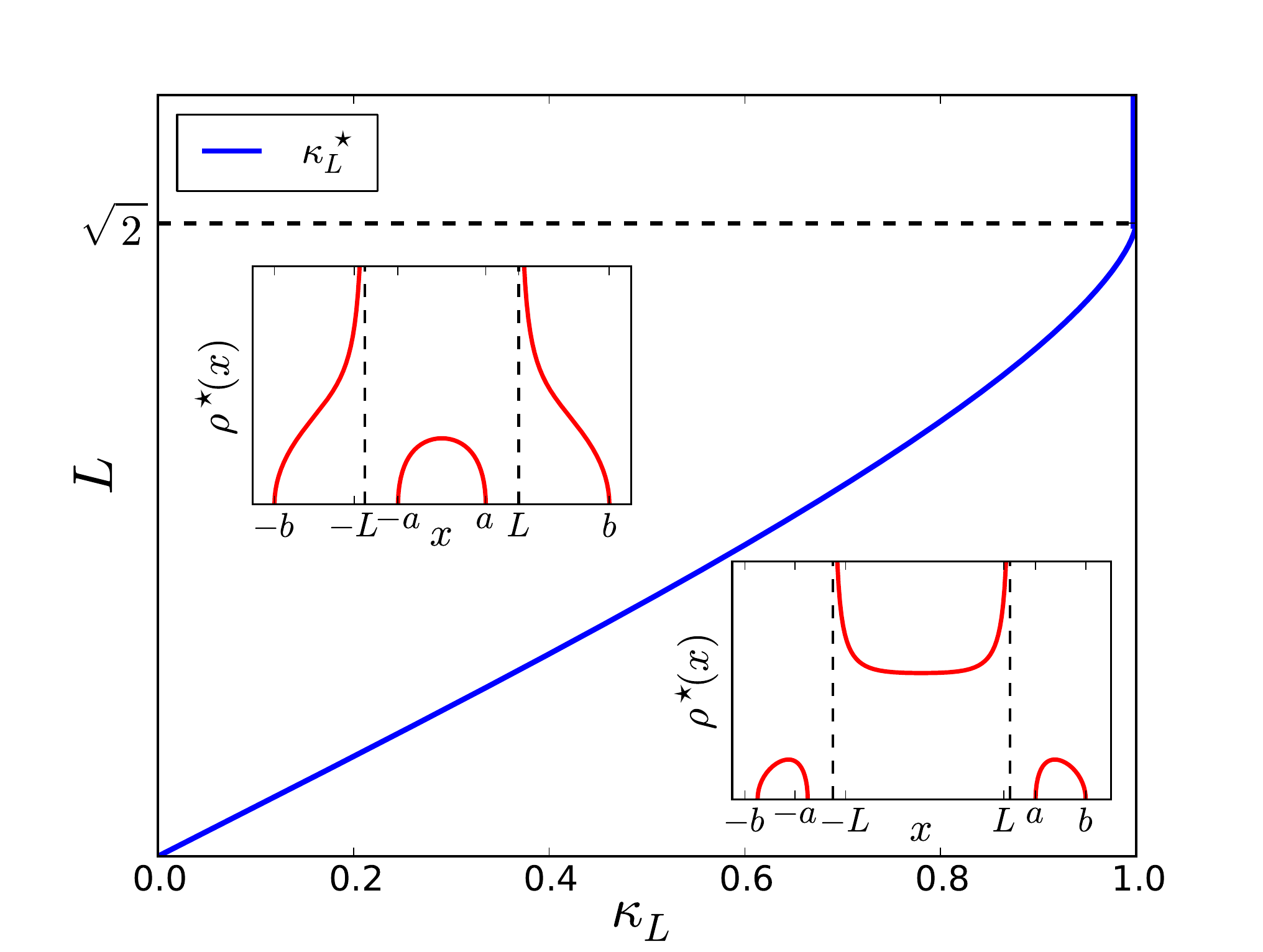} 
\caption{Phase diagram of the Coulomb gas (\ref{eq:hamilt_1}) in the $(\kappa_L, L)$ plane. The equilibrium density $\rho^\star(x)$, plotted in insets, exhibits a transition between two distinct shapes as the critical solid blue line $\kappa_L^\star$ is crossed, along which $\rho^\star$ is given by the Wigner's semi-circle density.}
\label{kappastar_L}
\end{figure}

\noindent where $\mathrm{Pr}$ denotes the Cauchy principal part, and $\mathrm{supp}(\rho^\star)$ the region on the real line where $\rho^\star(x)>0$. Eq. \eqref{eq:cauchy} is to be solved with the constraints $\int_{-L}^L \rho^\star(x)\dd x=\kappa_L$ and $\int_{-\infty}^\infty \rho^\star(x)\dd x=1$. This integral equation \eqref{eq:cauchy} can be solved using the resolvent method \cite{suppmat}. The normalized density has generally a three-cut support (see Fig. \ref{kappastar_L}), 
\begin{eqnarray}
\rho^\star(x)= \frac{1}{\pi}\sqrt{\frac{(x^2-b^2)(x^2-a^2)}{L^2-x^2}},
\label{f_of_x}
\end{eqnarray}
which is valid for $x$ belonging to any of the intervals in the support. The value of the edges $a$ and $b$ are determined by the relation 
$a^2+b^2=L^2+2$ together with the constraint $\int_{-L}^L \rho^\star(x)\dd x=\kappa_L$. 

For a fixed value of $L$, the equilibrium density will take three different shapes (see Fig. \ref{kappastar_L}) according to the fraction $\kappa_L$ of particles stacked in the box. If $\kappa_L=\kappa_L^\star$ (solid blue line in Fig. \ref{kappastar_L}), as many particles are stacked in the box as naturally expected from \eqref{gibbs} without any constraint. Thus the equilibrium density is just the semi-circle. For the cases $\kappa_L>\kappa_L^\star$ $(\kappa_L<\kappa_L^\star)$, an excess of particles accumulates inside (outside) the box, giving rise to three disconnected blobs and a divergence of the density $\rho^\star$ around the inner (outer) box walls (see Fig. \ref{kappastar_L}).

Evaluating \eqref{resulting} at the saddle point, we obtain a large deviation decay of the probability for large $N$ of the form in Eq. \eqref{generalrate}, where the rate function is given by
$
 \psi_L(\kappa_L)=\frac{1}{2}\left[S[\rho^\star]-S[\rho_{\mathrm{sc}}]\right] 
$
where $\rho_{\rm sc}$ is the Wigner's semi-circle density. The second term comes from the large $N$ behavior of the normalization constant $Z_{N,\beta}$ (\ref{jpd}) and needs to be subtracted. The rate function $\psi_L(\kappa_L)$ is therefore determined by the action \eqref{eq:hamilt_1} at the saddle point $S[\rho^\star]$. Its full expression is rather cumbersome (see \cite{suppmat}) but can be easily plotted as shown in Fig. \ref{rate_3D} for two different values of $L$. In order to extract the variance $V_N(L)$ from the rate function, we 
expand $\psi_L(\kappa_L)$ around its minimum $\kappa_L=\kappa_L^\star$. We first notice that $\psi_L(\kappa_L)$ has a different shape for $N^{-1} \ll L < \sqrt{2}$ (bulk) and $L>\sqrt{2}$ (tail), while the edge $L \sim \sqrt{2}$ needs a separate treatment.

{\it The bulk $N^{-1} \ll L < \sqrt{2}$.} In this case, $\psi_L(\kappa_L)$ is two-sided, and setting $\kappa_L=\kappa_L^\star-\delta$ \cite{suppmat}, we find that close to its minimum $\kappa_L^\star$, it behaves like
$
\psi_L(\kappa_L=\kappa_L^\star -\delta)\sim (\pi^2/4) \delta^2/\ln{(L(2-L^2)^{\frac{3}{2}}/|\delta|)}$ as $\delta\to 0$.
Therefore, around the critical value $\kappa_L^\star$ for ``sufficiently large" $L(2-L^2)^{\frac{3}{2}} \gg |\delta|$, the rate function is non-analytic and displays a quadratic behavior modulated by a logarithmic singularity. The physical origin of this non-analytic behavior is linked to a phase transition in the associated Coulomb gas when $\kappa_L$ crosses the critical value $\kappa_L^\star$ (see Fig. \ref{kappastar_L}). Inserting this behavior (close to $\kappa_L^\star$) into \eqref{generalrate}, using that $\delta = \kappa_L^*-\kappa_L = {\cal O}(1/N)$, we find that $\mathcal{P}_{\scriptscriptstyle N,L}(N_L)$ has a Gaussian behavior around $\kappa_L^\star N$, with a variance growing as in Eq. \eqref{variance_summary} (first line), thus recovering Dyson's bulk behavior \emph{away} from the edge. This Gaussian limiting distribution is thus valid on a scale $\sim\mathcal{O}\left(\sqrt{\ln(NL(2-L^2)^{\frac{3}{2}})}\right)$ around $\kappa^\star N$. However, beyond this scale, the fluctuations of $\kappa_L N$ 
are instead described by the full large deviation function in Eqs. \eqref{generalrate} which has non-Gaussian tails \cite{suppmat}. This analysis holds in the bulk, for a fixed $N^{-1} \ll L < \sqrt{2}$, but breaks down for $L \sim \sqrt{2}$ (edge) and in the tail, $L > \sqrt{2}$.

{\it The tail, $L > \sqrt{2}$.} In this regime the width of the box is much larger than the semicircle sea, $L>\sqrt{2}$, and $\kappa_L^\star $ freezes to the value $1$ (see Fig. \ref{kappastar_L}), since on average \emph{all} the eigenvalues are contained within the box. Therefore the rate function $\psi_L(\kappa_L)$ is one-sided. Setting $\kappa_L=1-\delta$ and expanding $\psi_L(\kappa_L)$ to leading order in $\delta>0$, one obtains a \emph{linear} behavior~\cite{suppmat}
$
\psi_L(\kappa_L=1-\delta)\sim \phi(L)\delta $, as $\delta\to 0$, 
where
\be
\phi(L)=L\sqrt{L^2-2}/2+\ln\left((L-\sqrt{L^2-2})/\sqrt{2}\right) \label{phiL} \;.
\ee
This function $\phi(L)$ turns out to be identical to the large deviation function describing the right tail of the top eigenvalue $\lambda_{\max}$, ${\rm Prob.} [\lambda_{\max} > L] \approx {\exp[-\beta N \phi(L)]}$. Here $N \phi(L)$ is the energy cost to pull out one particle at a distance larger than $L$ away from the Wigner sea, while the density of the rest $(N-1)$ charges remains of the standard semi-circular form~\cite{majumdar_vergassola_09}. To understand this connection with the right tail of $\lambda_{\max}$ one can extrapolate our Coulomb gas calculation of $\psi_L(\kappa_L)$ to the case where there is a discrete number of particles outside the interval $[-L, L]$. Because our Coulomb gas calculation preserves the symmetry $\rho^*(x) = \rho^*(-x)$, which is always true in the continuum limit, this extrapolation can only be done for an even number of particles, say two of them: one in the interval $(-\infty,-L]$ and the other one in the interval $[L, + \infty)$, hence in this case $\delta = 2/N$. Therefore, the energy cost of such a configuration, given by $\beta N^2 \psi_L(\kappa = 1-2/N)$, is precisely twice (as there are two particles) the energy $N \phi(L)$ to pull out one charge, at a distance $L>\sqrt{2}$, outside the Wigner sea (ignoring correlation effects between the two particles). Hence $N^2 \psi_L(\kappa = 1-2/N) \sim 2 N \phi(L)$,  in agreement with $
\psi_L(\kappa_L=1-\delta)\sim \phi(L)\delta $, as $\delta\to 0$. By using a similar energetic argument~\cite{majumdar_vergassola_09}, one can further show that $ \mathcal{P}_{\scriptscriptstyle N,L}(N_L = N-k) \sim A e^{-k N \beta \phi(L)}$, with $A = (1-e^{-N \beta \phi(L)})$~\cite{suppmat}, which is valid for $k \ll N$. The number variance $V_N(L)$ can be easily computed from this (discrete) exponential distribution, yielding $V_N(L) \sim e^{- N \beta \phi(L)}$, as announced in Eq.~(\ref{variance_summary}).

{\it The edge, $|L-\sqrt{2}| \sim {\cal O}(N^{-2/3})$}. This regime smoothly connects the other two ones. The number variance suddenly drops down to zero when $L$ approaches the edge of the semicircle $L\sim\sqrt{2}$. In this regime, the probability distribution $\mathcal{P}_{\scriptscriptstyle N,L}(N_L = N-k)$ can be expressed, for $\beta = 1, 2$ and 4 \cite{TW_GUE,TW_GOE}, in terms of Fredholm determinants, or equivalently as integrals involving a special solution of the Painlev\'e II equation. Computing the number variance from these expressions is however quite difficult. A simpler way to compute $V_N(L)$ in this regime is to resort to a finite $N$ calculation and then take the large $N$ limit in the edge scaling limit. We illustrate this approach in the case of GUE ($\beta=2$) -- but it could be extended to $\beta = 1$ and $4$. For $\beta = 2$, $V_N(L)$ is given by~\cite{mehta,forrester,fyodorov_lectures}
\be
V_N(L)=\int_{-L}^L \dd x K_N(x,x)- \int_{-L}^L \dd x \int_{-L}^L \dd y [K_N(x,y)]^2 \;.
\ee
Here $K_N(x,y)$ is the GUE kernel, expressed in terms of Hermite polynomials \cite{suppmat}.

We now zoom in the vicinity of the edge, setting $L=\sqrt{2}+{s}/(\sqrt{2}N^{2/3})$ \cite{forrester} and find that when $N \to \infty$, $V_N\left(\sqrt{2}+{s}/(\sqrt{2}N^{2/3})\right) \to \tilde V_2(s)$ where \cite{suppmat}
\be\label{variance_dble}
\tilde V_2(s)=2\int_s^\infty \dd x\ K_{\mathrm{Ai}}(x,x)-2\iint_{[s,\infty]^2}\dd x\dd y [K_{\mathrm{Ai}}(x,y)]^2 \;.
\ee
Here, $K_{\mathrm{Ai}}(x,y)$ is the Airy kernel given by $K_{\mathrm{Ai}}(x,y)=[\mathrm{Ai}(x) {\mathrm{Ai}}^\prime(y)-\mathrm{Ai}(y) {\mathrm{Ai}}^\prime(x)]/{(x-y)}$
where $\mathrm{Ai}(x)$ is the Airy function and, at coinciding points, $K_{\mathrm{Ai}}(x,x)=(\mathrm{Ai}^\prime (x))^2-x \mathrm{Ai}^2 (x)$. One can show that $\tilde V_2(s)$ behaves asymptotically as $\tilde V_2(s \to -\infty)\sim {3}/{(2\pi^2)} \ln |s|$ \cite{suppmat,Gus05} and $\tilde V_{2}(s \to \infty)\sim (8\pi)^{-1}s^{-\frac{3}{2}}e^{-\frac{4}{3}s^{{3}/{2}}}$ \cite{suppmat}. These asymptotic behaviors ensure a perfect matching with the behaviors of $V_N(L)$ on both sides of the edge for $L \gtrless \sqrt{2}$ in~\eqref{variance_summary}. For instance, when $L$ approaches $\sqrt{2}$ from above, $L \to \sqrt{2}^{+}$ one can substitute in the last line of (\ref{variance_summary}) the behavior of $\phi(L)$ when $L \to \sqrt{2}^+$, $\phi(L) \sim (2^{\frac{7}{4}}/3)(L-\sqrt{2})^{\frac{3}{2}}$. Hence, for $\beta = 2$, $V_N(L) \sim \exp{[-N (2^{\frac{11}{4}}/3)(L-\sqrt{2})^{\frac{3}{2}}]}$, which after a rearrangement of the argument coincides with the asymptotic behavior of $\tilde V_2(s \to \infty)$, with $s = \sqrt{2}N^{\frac{2}{3}}(L-\sqrt{2})$. We can similarly show that the matching also holds when $L$ approaches $\sqrt{2}$ from below. Assuming that this matching holds for all values of $\beta$, one expects the asymptotic behaviors
\begin{eqnarray}\label{asympt_vtilde}
\tilde V_\beta(s) \sim
\begin{cases}
&\frac{3}{\beta \pi^2} \ln |s| \;, \; s \to -\infty \\ 
&\exp{\left(-\frac{2 \beta}{3} s^{3/2}\right)} \;, \; s \to \infty \;.
\end{cases}
\end{eqnarray}
In Fig. \ref{fig_variance} we show a plot of $\tilde V_2(s)$ together with a comparison with numerical simulations, showing an excellent agreement. 

In conclusion, our exact large $N$ results in Eq. (\ref{variance_summary}) for GUE random matrix (with $\beta = 2$)
provides an analytic description of $V_N(L)$ for 1d spinless fermions in the
ground state, which was computed before only numerically \cite{vicari_pra,eisler_prl}. They may also be relevant to the statistics of entanglement entropy in such systems \cite{vicari_pra}. Even though most experiments in cold atoms are 
performed in bosonic systems (rather than fermionic), our results are expected to apply to bosonic systems in presence of
very strong repulsive interactions between bosons \cite{Lieb}.

We also expect a similar ``drop-off" effect (see Fig. \ref{fig_variance}) to occur in the statistics of the so-called \emph{index}, i.e. the number of positive eigenvalues in a given interval. The index is relevant for instance to describe the energy landscape of complex and glassy systems \cite{cavagna_garrahan_giardina_00,nadal_fyodorov} and was thus recently studied for different ensembles \cite{nadal_fyodorov,majumdar_nadal_scardicchio_vivo_09,majumdar_nadal_scardicchio_vivo_11,MMSV14}. Here, we have shown that the presence of the edge induces the ``drop-off'' effect. Hence it would be interesting to compute the number variance for random matrix ensembles without an edge, such as the Cauchy ensemble (see for instance \cite{MSVV13} and references therein).

\acknowledgments

 We acknowledge valuable correspondence with Viktor Eisler and Ettore Vicari. SNM and GS acknowledge support by ANR grant
2011-BS04-013-01 WALKMAT and in part by the Indo-French Centre for the Promotion of Advanced Research under Project 4604-3. GS and PV acknowledge support from Labex-PALM (Project Randmat).

\end{document}